Title:

**Accelerated calibrationless parallel transmit mapping using joint transmit and receive low-rank tensor completion**

Aaron.T. Hess[1*], Iulius Dragonu[2], Mark Chiew[3]

1. Oxford Centre for Clinical Magnetic Resonance Research (OCMR), University of Oxford
2. Siemens Healthcare Limited, Frimley, United Kingdom
3. Nuffield Department of Clinical Neurosciences, Wellcome Centre for Integrative Neuroimaging, FMRIB, University of Oxford, United Kingdom

Corresponding author: Dr Aaron Hess, Level 0, John Radcliffe hospital, Hedley way, Oxford, OX3 9DU, UK. aaron.hess@cardiov.ox.ac.uk

## Abstract

**Purpose**: To evaluate an algorithm for calibrationless parallel imaging to reconstruct undersampled parallel transmit field maps for the body and brain.

**Methods**: Using synthetic data, body, and brain measurements of relative transmit maps, three different approaches to a joint transmit-receive low-rank tensor completion algorithm are evaluated. These methods included: (i) virtual coils using the product of receive and transmit sensitivities, (ii) joint-receiver coils that enforces a low rank structure across receive coils of all transmit modes, and (iii) transmit low rank (TxLR) that uses a low rank structure for both receive and transmit modes simultaneously. The performance of each are investigated for different noise levels and different acceleration rates on an 8-channel parallel transmit 7T system.

**Results**: The virtual coils method broke down with increasing noise levels or acceleration rates greater than two producing RMS error greater than 0.1. The joint receiver coils method worked well up to acceleration factors of four, beyond which the RMS error exceeded 0.1. While TxLR enabled an eight-fold acceleration with most RMS errors remaining below 0.1.

**Conclusion**: This work demonstrates that under-sampling factors of up to eight-fold are feasible for transmit array mapping and can be reconstructed using calibrationless parallel imaging methods.

# 1. Introduction

Parallel transmit (pTx) technology mitigates transmit field heterogeneity(1,2), accelerates spatial RF pulses(3,4), and lowers SAR deposition(5–7), which is achieved by driving multiple (parallel) transmit coils with subject, target and pulse specific amplitudes and phases. RF pulses are designed by optimising a cost function using measured transmit fields. This work investigates the acceleration of pTx field mapping, with the goal of streamlining the use of pTx.

Universal pulses(8) have been demonstrated as an effective way of streamlining the use of pTx, however, for specific targets such as single voxel spectroscopy or anatomy that cannot be generalised including the heart, liver, prostate and spinal cord, subject specific field mapping with personalised RF pulse design is still required.

To map parallel transmit fields both absolute maps of B1+ and relative magnitude and phase can be measured. Relative maps are measured by transmitting in different transmit configurations, usually transmitting on one channel at a time while receiving on all receive channels all the time.

In the heart, for example, gating to the cardio-respiratory motion increases the measurement time by at least 5 fold for mapping transmit fields in the heart (assuming a respiratory efficiency of 50%, and cardiac acquisition window of 400 ms). Either multiple breathholds are required to cover the whole heart, or freebreathing using self-gating(9) can be used, the latter takes about 3 minutes, and the former in excess of 5 minutes. To reduce these measurement times, accelerated imaging techniques can be used to recover pTx field maps from under-sampled measurements. Accelerating the pTx field calibration will both increase the scan time available for diagnostic imaging and streamline the pTx adjustments.

Parallel transmit fields can be calculated using the same methodology as used for receive sensitivity mapping, and as demonstrated by Padormo et al, the same k-space calibration data can be used to estimate both receive sensitivities and transmit sensitivities (precise radiofrequency inference from multiple observations, PRIMO(10)). However, even when using a compact calibration region (24x24x24), it would take nearly 2 minutes to calibrate an 8 channel transmit system with respiratory and cardiac gating, which is significantly longer than a single breathhold.

In-order to recover missing k-space data calibrationless parallel imaging is required. One way calibrationless parallel imaging can be achieved using prior knowledge about the low-rank Hankel (or block-Hankel) structure of matrices formed from local k-space neighborhoods(11–14). The simultaneous autocalibrating and k-space estimation (SAKE)(13) method uses a projection-onto-sets algorithm with singular value thresholding to recover missing k-space data by alternating between enforcing data consistency and low-rank Hankel matrix structure, without the need of any fully-sampled calibration region. What methods like SAKE, and related methods like P-LORAKS(11,12) or ALOHA(15) exploit are redundancies that are present when Hankel matrices from multi-coil k-space data are combined, to produce a low-rank representation of the data which can be used to effectively constrains under-sampled data recovery. In fact, these approaches are also similar to methods such as ESPIRiT(16) and PRIMO(10), which form very similar low-rank Hankel-structured matrices, except that these latter methods work on fully-sampled data.

In this work we investigate the extension of these calibrationless parallel imaging approaches to recover undersampled relative transmit field maps using a structured low rank tensor completion. Here, we make use of the additional redundancy provided by a multi-dimensional data representation that has both transmit and receive field dimensions, by varying under-sampling patterns across different transmit encodings and effectively sharing information across these encodings. We investigate, in simulations and experimental data in the body and the brain, the acceleration factors achievable with this approach, and demonstrate that under-sampling factors of up to 8x are feasible with our proposed method.

## 2. Theory

### 2.1 Matrix structure

SAKE and the subspace identification part of ESPIRiT(16) take k-space data, and through the application of a kernel transform, form a block-Hankel representation. The methods differ in how the receive (Rx) channels are handled in the matrix structure, where SAKE vertically concatenates Rx coils, while ESPRiT horizontally concatenates Rx coils. Subspace identification is performed through low-rank reconstruction of missing samples (SAKE) or SVD-based singular value truncation (ESPIRiT).

PRIMO(10) extends ESPIRiT to include transmit (Tx) sensitivities, which is achieved by horizontally concatenating the Tx sensitives and vertically concatenating the Rx sensitives. The time interleaved acquisition of modes (TIAMO)(17) method harnesses transmit sensitives in the image reconstruction by using different transmit modes (sensitives) as virtual receive channels, this makes the number of virtual receiver channels a the product of receive channels and transmit modes (sensitives). We refer to these two methods below as PRIMO and virtual coil (VC).

Here we consider the VC and PRIMO representations to be special cases of a block-Hankel tensor formed from the multi-dimensional data $\boldsymbol{\mathcal{D}} \in \mathbb{C}^{N_{kx} \times N_{ky} \times N_{Rx} \times N_{Tx}}$, where the 4th order tensor (or multi-dimensional array) $\boldsymbol{\mathcal{D}}$ is transformed to another 4th order tensor $\boldsymbol{\mathcal{H}} \in \mathbb{C}^{N_1 \times N_2 \times N_{Rx} \times N_{Tx}}$ with block-Hankel frontal slices. The linear operator:

$$T: \mathbb{C}^{N_{kx} \times N_{ky} \times N_{Rx} \times N_{Tx}} \to \mathbb{C}^{N_1 \times N_2 \times N_{Rx} \times N_{Tx}} \tag{1}$$

is what maps the k-space data corresponding to any given Rx/Tx k-space into the block-Hankel $N_1 \times N_2$ matrix, by rastering over k-space with a kernel of dimension $N_1 = m \times n$ (e.g. $N_1 = 9$ for a $3 \times 3$ kernel), resulting in $N_2$ distinct kernel vectors. This is done for every Rx/Tx pair, which results in the tensor $\mathcal{H} = T(\mathcal{D})$ as demonstrated in Figure 1.

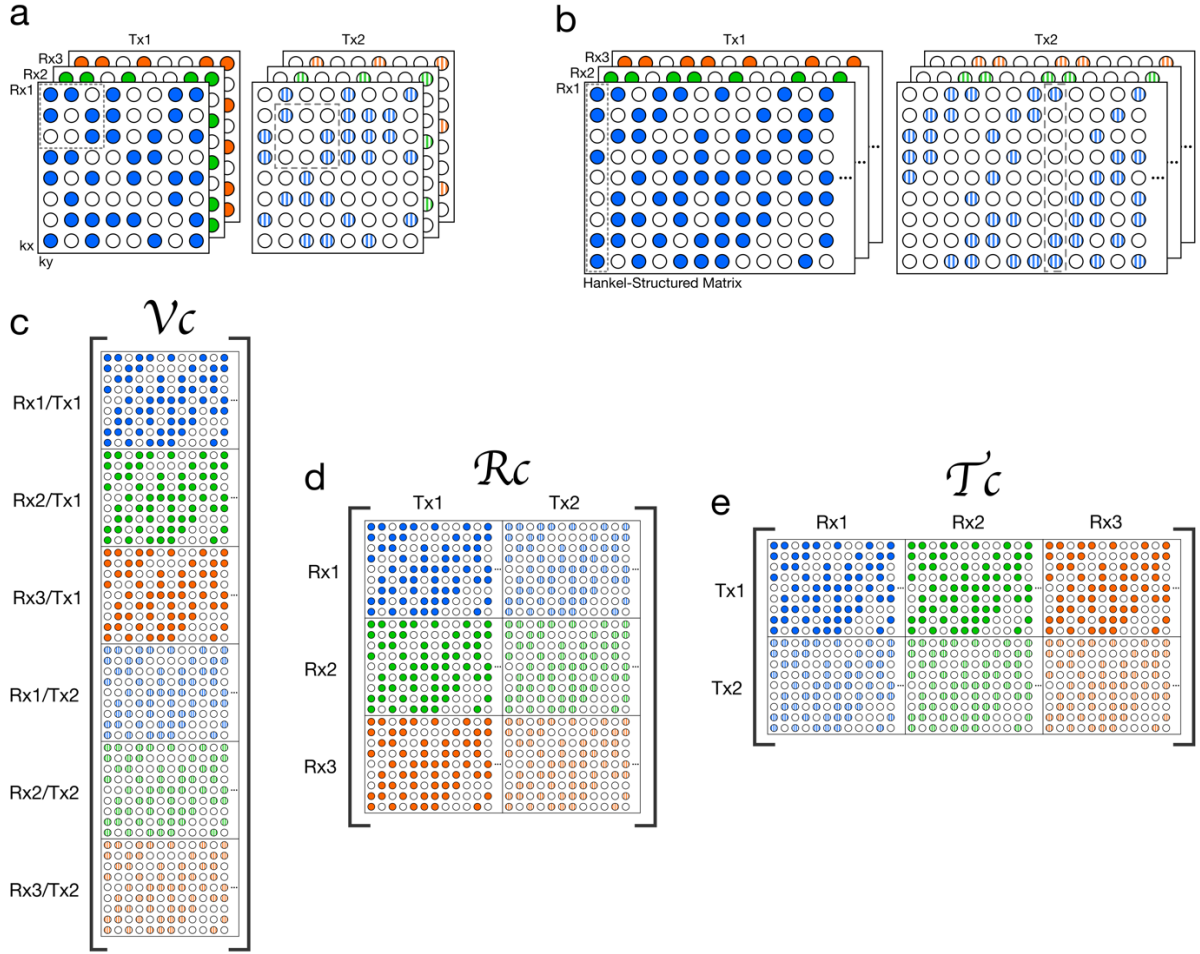

Figure 1 - Diagram showing how (a) a multi-dimensional k-space dataset with dimensions kx, ky, Rx and Tx can be transformed to (b) a 4$^{th}$ order tensor with block-Hankel frontal slices. (c-e) show different matricizations or unfoldings of the tensor, corresponding to: (c) the virtual coil matrix ($\mathcal{V}_C$), (d) the PRIMO-style receive vertical concatenation matrix ($\mathcal{R}_C$), and (e) the transmit vertical concatenation matrix ($\mathcal{T}_C$). The proposed TxLR method uses low-rank constraints on both $\mathcal{R}_C$ and $\mathcal{T}_C$ unfoldings of the tensor.

There is some freedom in the way these multi-dimensional k-space data are transformed into matrices (i.e. matricization), particularly in the axis along which the receive (and transmit) channels are concatenated. This choice, of vertical or horizontal concatenation was recently investigated by Jin et al(18), where they found that horizontal concatenation across Rx channels to produce better reconstructions than vertical concatenation (as used in SAKE).

Virtual coils (VC): From the tensor $\mathcal{H}$, we can construct the matrix corresponding to the interpretation of every Rx/Tx pair as a unique receive map by applying a particular unfolding of the tensor

$$U_0: \mathbb{C}^{N_1 \times N_2 \times N_{Rx} \times N_{Tx}} \rightarrow \mathbb{C}^{(N_1 \cdot N_{Rx} \cdot N_{Tx}) \times (N_2)} \tag{2}$$

, such that $\mathcal{V}_C = U_0(\mathcal{H})$ (Fig. 1c).

Receive concatenation: Alternatively we can construct a matrix corresponding to the *calibration matrix* in the PRIMO relative transmit mapping approach by applying the following unfolding of the tensor

$$U_2: \mathbb{C}^{N_1 \times N_2 \times N_{Rx} \times N_{Tx}} \rightarrow \mathbb{C}^{(N_1 \cdot N_{Rx}) \times (N_2 \cdot N_{Tx})} \tag{3}$$

, such that $\mathcal{R}_C = U_2(\mathcal{H})$ (Fig. 1d).

Transmit concatenation: We recognize here that the role to the Tx sensitivities is entirely analogous to that of the Rx sensitivities, and that if $\mathcal{R}_C$ should be well characterised with a low-dimensional subspace (i.e. it has low-rank), then we should also be able to form a symmetrical unfolding:

$$U_1: \mathbb{C}^{N_1 \times N_2 \times N_{Rx} \times N_{Tx}} \rightarrow \mathbb{C}^{(N_1 \cdot N_{Tx}) \times (N_2 \cdot N_{Rx})} \tag{4}$$

, such that the matrix $\mathcal{T}_C = U_1(\mathcal{H})$ (Fig. 1e) also has low-rank. All tensor unfoldings can be achieved through permutation and reshaping of the multi-dimensional array. However, we note that $\mathcal{T}_C$ is not simply the transpose of $\mathcal{R}_C$, due to the fact that the block-Hankel front slices of the tensor are not changed when the Rx and Tx dimensions are swapped.

One key idea that we explore here is not only the alternative choice of Rx and Tx concatenations (compared to the PRIMO-style concatenation), but also the use of both representations and the simultaneous enforcement of low-rank structure in both $\mathcal{T}_C$ and $\mathcal{R}_C$ unfoldings. We propose to use this structure of the Rx/Tx k-space to exploit all the correlations and redundancies in the multi-dimensional data, by reconstructing under-sampled k-space Rx/Tx data through the use of low-rank constraints across *multiple* simultaneous unfoldings of the data tensor and taking advantage of the symmetry between the Rx and Tx dimensions.

## 2.2 Low-Rank Constraints

We can formulate the reconstruction problem as a convex problem using a sum-of-nuclear norms approach as follows:

$$\min_{\boldsymbol{z}} \frac{1}{2}\|M\boldsymbol{z} - \boldsymbol{D}\|_2^2 + \lambda_1 \|U_1(T(\boldsymbol{z}))\|_* + \lambda_2 \|U_2(T(\boldsymbol{z}))\|_* \tag{5}$$

However, in this work we choose to employ an alternative formulation with strict rank constraints, that while non-convex, performs better than the convex formulation and has the advantage of a more intuitive constraint parameterization (rank thresholds, rather than $\lambda$):

$$\min_{\boldsymbol{z}} \frac{1}{2}\|M\boldsymbol{z} - \boldsymbol{D}\|_2^2 \tag{6}$$

$$subject\ to: rank(\mathcal{T}_{\mathcal{C}}) = r_1$$

$$subject\ to: rank(\mathcal{R}_{\mathcal{C}}) = r_2$$

Here, $M$ is the k-space sampling mask (applied to $\boldsymbol{z}$ an element-wise Hadamard product), $\boldsymbol{z}$ is the reconstructed k-space tensor, $\boldsymbol{D}$ is the under-sampled k-space data, $\mathcal{T}_{\mathcal{C}} = U_1(T(\boldsymbol{z}))$, $\mathcal{R}_{\mathcal{C}} = U_2(T(\boldsymbol{z}))$, and $r_i$ are the rank thresholds. We will refer to this method for accelerated relative parallel transmit mapping as the transmit low rank (TxLR) method, which we can also reference using the constraint shorthand $\mathcal{T}_{\mathcal{C}} + \mathcal{R}_{\mathcal{C}}$.

Using the same reconstruction framework, with only changes to the number of constraints and the type of unfolding operator used, we can formulate a "VC" reconstruction that is the analogue of the virtual coil (TIAMO or VC) approach:

$$\min_{\boldsymbol{z}} \frac{1}{2}\|M\boldsymbol{z} - \boldsymbol{D}\|_2^2 \tag{7}$$

$$subject\ to: rank(\mathcal{V}_{\mathcal{C}}) = r_0$$

where $\mathcal{V}_{\mathcal{C}} = U_0(T(\boldsymbol{z}))$.

We will refer to the $\mathcal{R}_\mathcal{C}$ constrained reconstruction as the "PRIMO" approach, due to the correspondence between the transmit and receive concatenation order in $\mathcal{R}_\mathcal{C}$ and the relative transmit mapping calibration matrix defined by PRIMO. The PRIMO reconstruction is formulated as:

$$\min_{\boldsymbol{z}} \tfrac{1}{2}\|M\boldsymbol{z} - \mathcal{D}\|_2^2$$
$$\text{subject to}: rank(\mathcal{R}_\mathcal{C}) = r_2 \tag{8}$$

where $\mathcal{R}_\mathcal{C} = U_2(T(\boldsymbol{z}))$. The difference between the proposed TxLR approach and the PRIMO approach is the additional simultaneously enforced $\mathcal{T}_\mathcal{C}$ constraint in the TxLR formulation.

## 3. Methods

### 3.1 Algorithm

The constrained optimisation problem in Eq. 6 was solved using the alternating direction method of multipliers (ADMM)(19) algorithm:

$$\begin{aligned}
(i):\ & \mathcal{T}_\mathcal{C}^{n+1} = \Gamma_{r_1}\left(U_1(T(\boldsymbol{z}^n - \boldsymbol{y}_1^n))\right) \\
(ii):\ & \mathcal{R}_\mathcal{C}^{n+1} = \Gamma_{r_2}\left(U_2(T(\boldsymbol{z}^n - \boldsymbol{y}_2^n))\right) \\
(iii):\ & \boldsymbol{z}^{n+1} = argmin\ \|M\boldsymbol{z} - \mathcal{D}\|_2^2 \\
& \quad + \tfrac{\rho}{2}\left\|\mathcal{T}_\mathcal{C}^{n+1} - U_1(T(\boldsymbol{z})) + \boldsymbol{y}_1^n + \mathcal{R}_\mathcal{C}^{n+1} - U_2(T(\boldsymbol{z})) + \boldsymbol{y}_2^n\right\|_2^2 \\
(iv):\ & \boldsymbol{y}_1^{n+1} = \boldsymbol{y}_1^n + \mathcal{T}_\mathcal{C}^{n+1} - U_1(T(\boldsymbol{z}^{n+1})) \\
(v):\ & \boldsymbol{y}_2^{n+1} = \boldsymbol{y}_2^n + \mathcal{R}_\mathcal{C}^{n+1} - U_2(T(\boldsymbol{z}^{n+1}))
\end{aligned} \tag{9}$$

where superscript denotes the iteration number, and $\boldsymbol{y}_1$ and $\boldsymbol{y}_2$ are auxiliary variables in the ADMM algorithm. The final output $\boldsymbol{z}^n$ after $n$ iterations is a k-space by Rx by Tx tensor with the same dimensions as the under-sampled $\mathcal{D}$. $\Gamma_r$ is a singular value hard-thresholding operator that performs an SVD and discards all singular values with index $i > r$, when ordered in decreasing magnitude. The algorithm is initialised with all variables, including $\boldsymbol{z}^0$, set to $\boldsymbol{0}$ (a matrix or array of zeros).

The least squares optimisation in line ($iii$) of the algorithm is performed analytically in a single step. To do this, the adjoint operators $T^*, U_i^*$ need to be defined. $T^*$, the adjoint block-Hankel operator, takes each column vector of the block-Hankel structured tensor, reshapes that into a 2D array, and places it back into its corresponding k-space location. Overlapping values are simply *summed* in place. The pseudo-inverse operator $T^\dagger$, which is not used here, instead *averages* overlapping values. $U_i^*$, the adjoint unfolding operators are simply re-folding (reshaping and permuting) the matrix back into the original multi-dimensional array shape.

Reconstructions were run to a fixed number of iterations, or using a chi-square heuristic based on receive array noise characteristics. The chi-square heuristic checks, at every iteration, whether the value:

$$\sum_{i=1}^{N_{Rx}} \frac{\|Mz_i - \mathcal{D}_i\|_2^2}{\sigma_i^2} \Big/ v > 1 \tag{10}$$

where $\sigma_i$ is the noise variance from the $i^{th}$ Rx channel (obtained from a separately acquired noise reference scan), and $v$ is the total number of sampled data points. This expression is the well known chi-square goodness of fit test, and we use it here to provide a parameter-free stopping criterion.

All reconstructions were implemented using MATLAB R2019b (Mathworks, Natick, MA, USA). The VC (Eq. 7) and PRIMO (Eq. 8) reconstructions were performed analogously, by omitting steps ($ii$) and ($v$), adjusting the second term in ($iii$) accordingly, and using the appropriate unfolding operator $U_0$ or $U_2$. In practice, the algorithm also uses a varying penalty parameter $\rho$, with $\rho^0 = 10^{-6}$ and scaling factor $\tau = 1.1$, and takes advantage of over-relaxation with parameter $\alpha = 1.5$ to improve convergence (see Boyd et al., 2010 for details).

### 3.2 Data

In-silico data were simulated in Sim4Life (ZMT, Zurich, Switzerland) using an eight channel transmit-receive dipole array(20) at the 7T frequency of 298 MHz, centred over the heart of Duke(21) (Virtual population, iTIS foundation, Zurich, Switzerland). The maximum resolution around the conductors was set to 0.5 mm, 1.0 or 2.0 as required to capture the geometry. A synthetic proton density was generated using the tissue density, with densities greater than 1200 kg/m³ (bone) or less than 400 kg/m³ (lung) set to 80 to make them resemble an MR image. The field of view was 278 x 356 x 248 mm³ (AP/LR/HF) data were re-sampled onto a uniform grid of 2x2x2 mm³. This data was Fourier transformed to form a k-space dataset, and 48 central slices located in the body were used for reconstruction simulations.

In-vivo data were acquired on a 7T Magnetom (Siemens, Erlangen, Germany), and all volunteers provided written informed consent and were scanned in accordance with our institutions ethical practices. Data were acquired using a spoiled low flip angle gradient echo acquisition which transmits on one channel at a time while receiving on all channels (see Supporting Information Table S1 for details of the protocols used). In the brain, data was acquired in a sagittal orientation using an 8 transmit, 32 Receive coil (Nova medical, Wilmington, USA) and a 3D acquisition. In the body an 8 channel transmit/receive dipole array was used(20) and 2D images were acquired in a horizontal long axis (HLA) orientation. Receive array noise measurements were also collected to generate noise covariance matrices for the chi-square convergence heuristic.

All simulations using the synthetic data were performed with added channel-independent noise at peak SNR (PSNR, defined as the maximum signal divided by the noise standard deviation) in k-space was 60 dB unless otherwise noted (i.e. the ratio of the maximum k-space magnitude across channels to the noise standard deviation was 60 dB). Retrospective under-sampling using a uniform density pseudo-random Poisson disc distribution was employed to assess reconstruction performance at acceleration factors ranging from $R = 2$ to $R = 12$

### 3.3 Code and Data Availability

All data are provided for reproducibility at the following location https://doi.org/10.5287/bodleian:o011mPAQB. Reconstruction code, and figure-generating code are also made fully available at https://users.fmrib.ox.ac.uk/~mchiew/research.html.

### 3.4 Reconstruction Parameters

Reconstructions were performed using a [5,5] kernel, and rank thresholds were set to 50 unless otherwise stated. The number of iterations was set to 50 for the TxLR reconstructions, and 100 for the VC and PRIMO methods in all synthetic data reconstructions, except in the cases where the chi-square heuristic was evaluated. See Supporting Figure S1 for validation of these parameter choices. The chi-square stopping criteria was used in all reconstructions of the in vivo body and brain data.

In all cases, the k-space data was cropped to a matrix size of 24x24, except where the impact of k-space matrix size was explicitly evaluated, which examined crop sizes of 24x24, 36x36 and 48x48. For the 3D datasets (synthetic body, and in vivo brain), reconstruction was performed slice-by-slice on hybrid space kx-ky-z (synthetic body) or x-ky-kz (brain) data.

Reconstruction fidelity was evaluated using a normalise root mean square error (RMSE) metric:

$$RMSE = \frac{\|\hat{z} - z\|_2}{\|z\|_2} \quad (11)$$

where $\hat{z}$ is the estimated k-space tensor, and $z$ is the ground truth. Here, the L2 norm is evaluated after vectorising the tensors. Relative transmit sensitivities were also estimated using the PRIMO extension to ESPIRiT after under-sampled tensor reconstruction.

## 4. Results

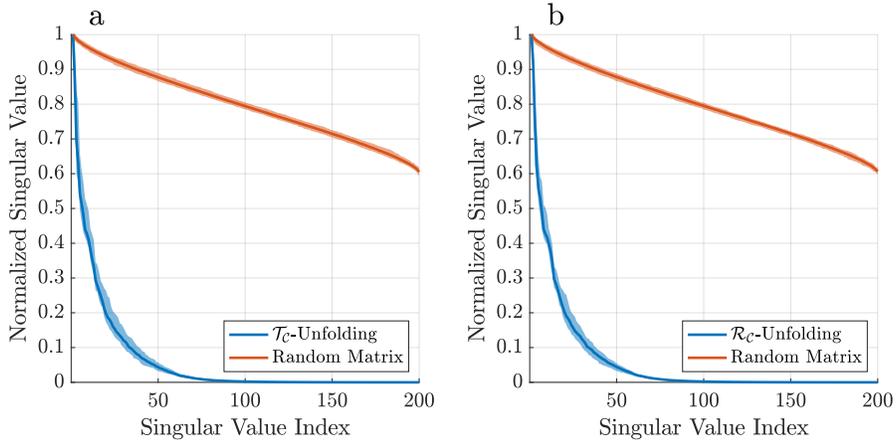

*Figure 2 - Singular value distributions for the $\mathcal{T}_\mathcal{C}$ (a) and $\mathcal{R}_\mathcal{C}$ (b) unfoldings of the synthetic data (blue). The shaded regions show the minimum and maximum range for the singular values across all slices in the dataset. In orange, singular value distributions for random matrices are shown for comparison.*

Figure 2 illustrates the singular value distributions across the $\mathcal{T}_\mathcal{C}$ and $\mathcal{R}_\mathcal{C}$ unfoldings in the synthetic dataset, compared to the singular values associated with random matrices of the same dimensions. The singular values for both distributions show similar low-rank structure, and suggest that exploitation of both unfoldings simultaneously can more effectively constrain the under-sampled k-space recovery.

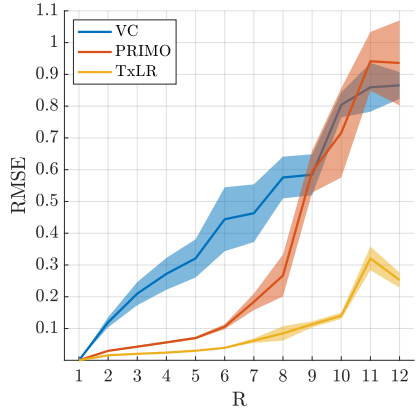

*Figure 3 - RMSE for the VC (blue), PRIMO (orange) and TxLR (yellow) methods evaluated in the synthetic body dataset. Shaded regions indicate the standard deviation of the RMSE across 48 slices.*

In Figure 3, reconstruction performance on the synthetic body dataset is shown for acceleration factors ranging from $R = 2$ to $R = 12$, for the VC approach ($\mathcal{V}_\mathcal{C}$ constraint), PRIMO approach ($\mathcal{R}_\mathcal{C}$ constraint), and the proposed TxLR approach ($\mathcal{T}_\mathcal{C} + \mathcal{R}_\mathcal{C}$ constraints). Mean RMSEs for all methods are shown, along with their standard deviations across all 48 reconstructed slices. While the VC reconstruction reaches RMSE > 0.1 at $R = 2$, the PRIMO reconstruction is able to achieve $R = 6$ under-sampling before reaching that error threshold. In comparison, the TxLR reconstruction manages to reach acceleration factors of $R = 8$ before showing the same error levels. Furthermore, both the VC and PRIMO reconstructions suffer from very poor reconstruction fidelity as acceleration factors increase, whereas the proposed TxLR approach shows a much lower rate of error increase with increasing $R$.

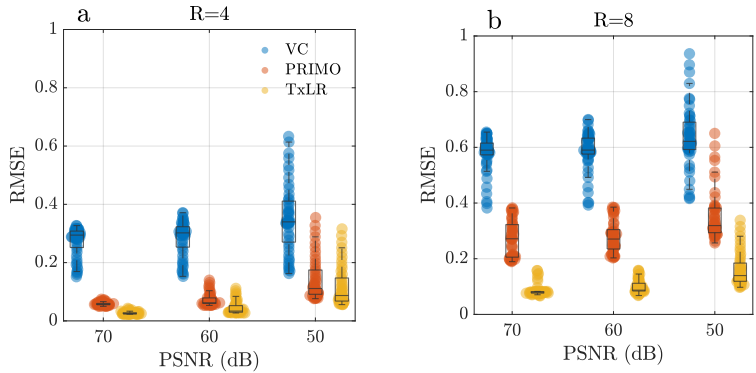

*Figure 4 - RMSE in the synthetic data at varying PSNR and acceleration factors. Each VC (blue), PRIMO (orange), and TxLR (yellow) scatter point represents the RSME for a single slice. Acceleration factors (a) R=4, and (b) R=8 are shown.*

Reconstructions in the synthetic dataset were assessed at varying noise levels, with PSNR ranging from 50 dB to 70 dB, across acceleration factors of $R = 4$ and $R = 8$. Figure 4 plots the aggregated RMSE from all 48 slices, which shows the expected increase in RMSE with decreasing PSNR, and increasing acceleration factors, across all methods. In all cases, the proposed TxLR reconstructions produce the lowest RMSE, although at the lower acceleration factor PRIMO performs nearly as well.

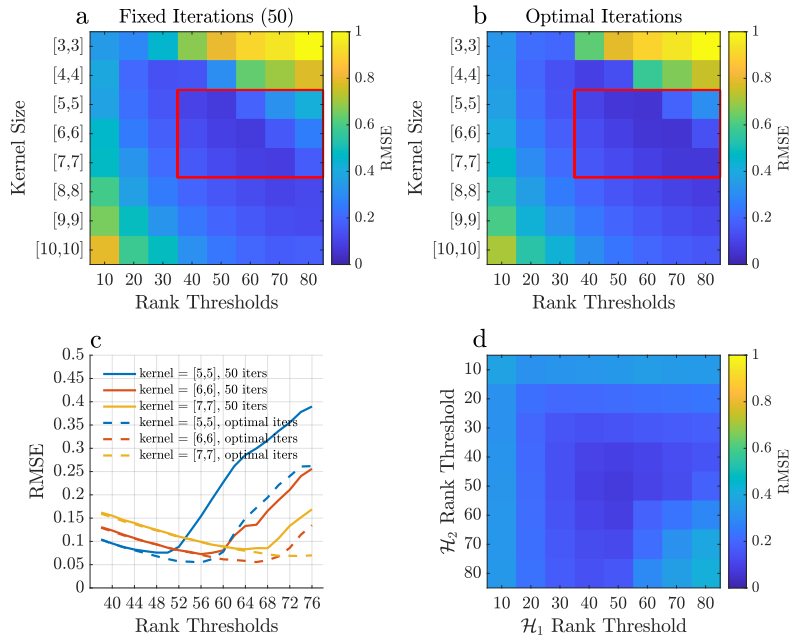

*Figure 5 - Exploring the impact of various parameter choices on the TxLR approach. (a) RMSE at different kernel sizes and rank thresholds, using a fixed 50 iteration reconstruction; (b) similar to (a), but with a retrospectively selected number of iterations that produces the smallest RMSE; (c) line plot of the parameter space outlined in the red boxes of (a,b); (d) the result of varying the $\mathcal{T_C}$ and $\mathcal{R_C}$ constraints independently.*

In Figure 5 the impact of varying iterations, kernel size and rank thresholds on the proposed reconstruction scheme was evaluated on a central slice of the synthetic dataset at R=8. Figures 5a and 5b show the reconstructed RMSE across square kernels ranging from [3,3] to [10,10] and rank thresholds of 10 - 80 for both the $\mathcal{T_C}$ and $\mathcal{R_C}$ constraints. In Fig. 5a, a fixed iteration count of 50 was used, whereas in Fig. 5b, the "optimal" number of iterations was used, based on retrospectively choosing the number of iterations that minimized RMSE. In both cases, the minima can be found within the parameter subset highlighted by the red box, for which Fig. 5c provides a more detailed comparison of RMSE performance. This highlights that a rank constraint of 50, in a fixed 50 iteration reconstruction produces near-optimal results. Although slight improvements can be found with different thresholds, this requires foreknowledge of the optimal iteration count, which is not typically available in a prospectively under-sampled acquisition. Fig. 5d shows the impact of varying the rank constraints of the $\mathcal{T}_\mathcal{C}$ and $\mathcal{R}_\mathcal{C}$ terms independently, indicating that RMSE is minimized when both rank constraints are the same and equal to 50 in this data.

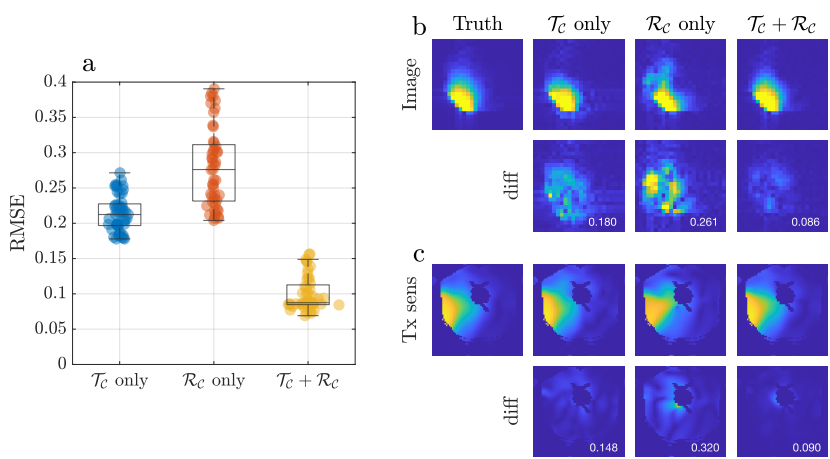

*Figure 6 - (a) RMSE in the synthetic data at R=8, resulting from reconstructions using the $\mathcal{T}_\mathcal{C}$ constraint (blue), $\mathcal{R}_\mathcal{C}$ constraint (orange, PRIMO), and simultaneous $\mathcal{T}_\mathcal{C} + \mathcal{R}_\mathcal{C}$ constraints (yellow, TxLR), with each marker corresponding to a single slice. (b) Reconstructed images and (c) transmit field maps corresponding to a representative Tx/Rx pair (Tx channel 4, Rx channel 3), at slice z=24. The number in the difference images correspond to RMSE.*

To further assess the value of the proposed approach, Figure 6 shows a comparison of using the $\mathcal{T}_\mathcal{C}$ or $\mathcal{R}_\mathcal{C}$ constraints individually, or simultaneously as proposed, evaluated again in the synthetic body dataset at R=8. In Figure 6a, RMSE is shown to be similar for the $\mathcal{T}_\mathcal{C}$ or $\mathcal{R}_\mathcal{C}$ only reconstructions (where $\mathcal{R}_\mathcal{C}$ only is the same as the PRIMO approach), with mean RMSE between 0.2 and 0.3. This result shows that while in this case, vertical concatenation of the Tx dimension (to form the $\mathcal{T}_\mathcal{C}$ matrix) slightly outperforms the $\mathcal{R}_\mathcal{C}$-based PRIMO approach of vertically concatenating the Rx dimension, the difference between either choice of Tx/Rx concatenation ordering is relatively small. However, reconstruction fidelity is significantly better for the TxLR approach which uses the simultaneous $\mathcal{T_C} + \mathcal{R_C}$ constraints, with mean RMSE below 0.1.

In Figure 6b, the top row shows a representative reconstructed image is shown for all three approaches compared to the ground truth. The second row shows difference images, along with the RMSE for that image, which is lowest in the $\mathcal{T_C} + \mathcal{R_C}$ case. The bottom two rows show an estimated relative transmit sensitivity map and the difference image from the ground truth. Again, both visual qualitative inspection and RMSE values indicate that the proposed $\mathcal{T_C} + \mathcal{R_C}$ (TxLR) reconstruction produces the lowest transmit field map errors.

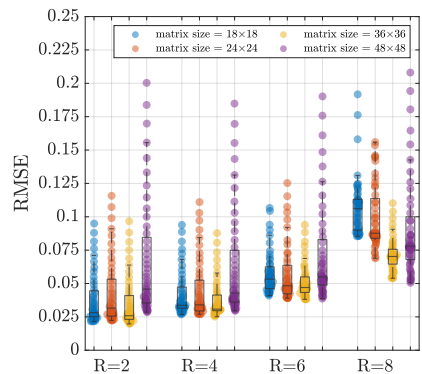

*Figure 7 - RMSE for different matrix sizes in the proposed method, at R=2, 4, 6 and 8. Each marker represents the RMSE from a single slice. Matrix sizes of 18x18 (blue), 24x24 (orange), 36x36 (yellow) and 48x48 (purple) were evaluated.*

Figure 7 shows, in the synthetic body dataset, the effect of different k-space matrix sizes across different acceleration factors in the proposed approach. Interestingly, it appears that RMSE is minimized when the k-space matrix size is 36x36, across all acceleration factors. However, these differences are mostly small, and the changes in RMSE are largely driven by acceleration factor. The largest matrix size of 48x48 performs worst at all acceleration factors except R=8, where the smaller matrix sizes of 18x18 and 24x24 had the highest mean RMSE.

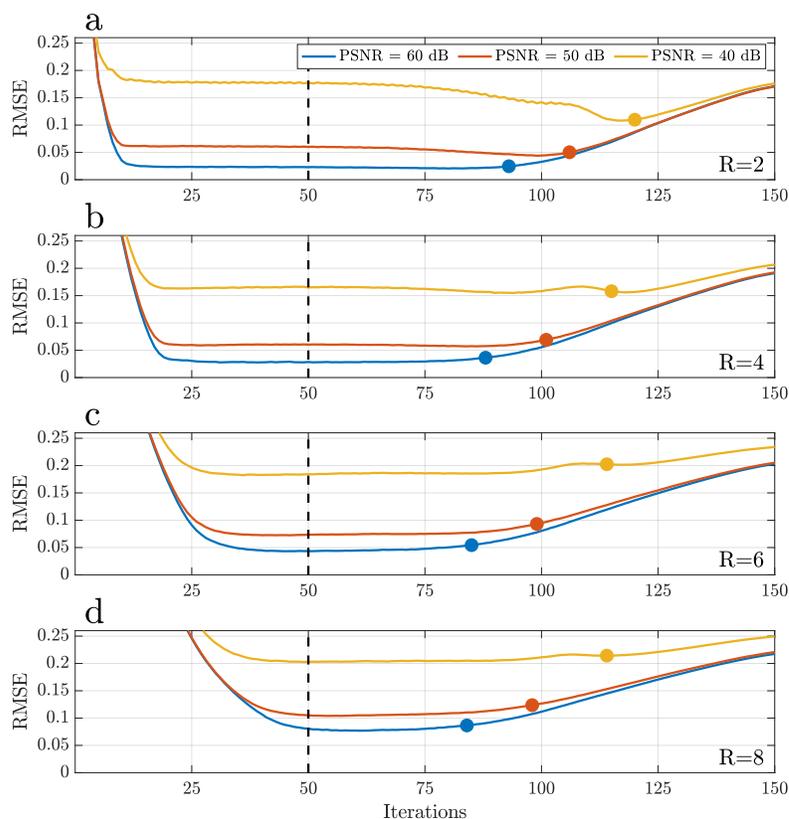

*Figure 8 - RMSE vs iteration count for the proposed method, evaluated at PSNRs of 40 (yellow), 50 (orange) and 60 dB (blue) and acceleration factors of (a) R=2, (b) R=4, (c) R=6, and (d) R=8. The RMSE at a fixed iteration count of 50 is denoted by the dashed black line, whereas the circles denote the number of iterations selected by the chi-square heuristic.*

An evaluation of the chi-square stopping heuristic is shown in Figure 8. The RMSE produced by a fixed iteration cutoff of 50 is compared to stopping iterations based on the chi-square heuristic, denoted by the circle markers. This was evaluated in the synthetic data on a central slice (z=24), at different PSNR levels (40 dB to 60 dB) and acceleration factors (R=2-8). We see that in these data, the choice to cut off iterations at 50 typically coincided with the beginning or middle of a plateau of the RMSE in all cases. In contrast, the chi-square heuristic typically selected for iteration parameters that corresponded the end of the plateau, in all cases, just before the RMSE begins to increase again as the reconstruction steps away from the local minimum. In either case, both fixed iterations and the chi-square heuristic were able to produce reconstructions very close to the local optimum, but the chi-square heuristic requires no *a priori* information (aside from receiver noise measurements) or guesswork.

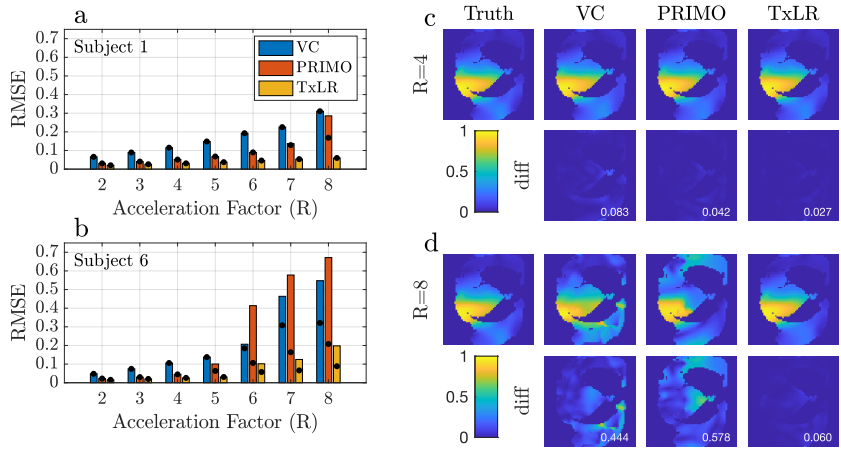

*Figure 9 - Results for the in vivo 2D body data. (a) RMSE for the best performing subject and (b) the worst performing subject, showing the VC (blue), PRIMO (orange) and TxLR (yellow) methods against acceleration factors of R=2-8. (c) Estimated transmit sensitivity maps and the magnitude of the complex differences to the ground truth for Subject 4, Tx channel 3, at R=4 and (d) R=8. The values in each difference map correspond to the sensitivity map complex RMSE compared to ground truth.*

Figure 9 shows the reconstruction results for the in vivo 2D body datasets. In Fig. 9a, results for the best (subject 1) and worst (subject 6) performing reconstructions are plotted (plots for all subjects can be found in the Supporting Figure S4). The bar chart values represent the RMSE at a rank constraint of 50, compared to an *optimal* rank constraint between 5-50 shown in the black dot. For the best performing subject, RMSE remained below 0.1 for the proposed TxLR method even at R=8 acceleration, whereas in the worst performing subject, RMSE reached as high as 0.2 at R=8. However, in all subjects and all acceleration factors, the TxLR reconstructions produced lower RMSE than the VC or PRIMO methods, although the differences between TxLR and PRIMO were relatively small for R≤4. In Subjects 1 and 2, the chosen rank constraint of 50 was optimal or close to optimal, which may reflect resolution and bandwidth differences in those datasets from compared to the others. In Subjects 3-6 the optimal rank constraint produced considerable lower RMSE for all methods, particularly in Subject 6.

In Figure 9b, we see relative Tx sensitivity maps estimated from ESPIRiT after tensor completion, and their difference images at R=4 and R=8, for a representative subject using the rank 50 reconstructions. Although the differences between methods are small at R=4, consistent with the results in the synthetic data, at R=8, only the TxLR method produces a Tx-sensitivity map that resembles the ground truth, with comparable error to the R=4 reconstruction.
1717

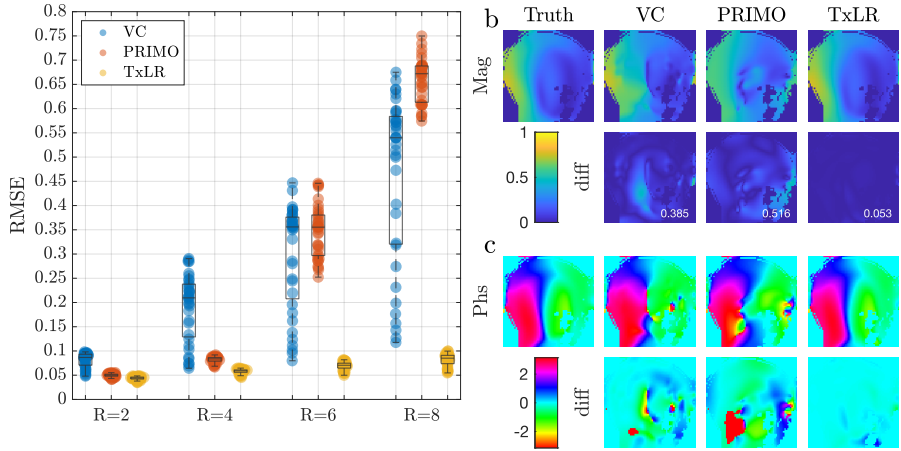

*Figure 10 - Results from the in vivo brain dataset. (a) RMSE at acceleration factors R=2, 4, 6 and 8, for the VC method (blue), PRIMO (orange) and TxLR (yellow), with markers representing the RMSE for each of the 36 slices. (b,c) Representative (Tx channel 4) transmit sensitivity maps, showing (b) magnitudes maps and magnitude differences, and (c) phase maps and phase difference images. The values inset on the difference maps in (b) correspond to the complex RMSE compared to ground truth*

Figure 10 shows the reconstruction results for the 3D in vivo brain dataset. In Fig. 10a, the RMSE across 36 central slices is shown, with the TxLR data showing a dramatic improvement in RMSE compared to the VC or PRIMO reconstructions at higher acceleration factors (R=6, 8), which could reflect better use of the increased redundancy provided by the 32-channel receive array (compared to the 8-channel receive used in the synthetic and experimental body data). Even at R=8, all brain slices were below RMSE of 0.1 (min=0.0547, max=0.0997). In contrast, the VC results show significantly increased RMSE beyond R=2, and PRIMO beyond R=4.

Figs. 10b and 10c show the magnitude and phase of relative Tx sensitivity maps and their difference images from the ground truth for Tx channel 4, for the R=8 data. The maps derived from the TxLR reconstructions show clear benefit over the other approaches, and are virtually indistinguishable from the ground truth, in both magnitude and phase. From the difference maps, it appears that errors in the TxLR maps are concentrated largely in the neck.

Reconstruction times (measured using an Intel Core i9, 8-core 2.3 GHz processor) in the body dataset were 0.18 s/iteration/slice (VC), 0.16 s/iteration/slice (PRIMO) and 0.20 s/iteration/slice (TxLR), for overall reconstruction times of approximately 10 - 20 s/slice. In the brain dataset, with 32 receive channels instead of 8, reconstruction times were 0.7 s/iteration/slice (VC, PRIMO) and 1.0 s/iteration/slice (TxLR) for total reconstruction times of approximately 50 - 70 s/slice.

## 5. Discussion

This work investigated the use of Hankel-structured low rank tensor completion to reconstruct under sampled parallel transmit field maps. This leverages the fact that both transmit and receive sensitivity maps possess low rank properties when represented as a Hankel-structured matrix. Redundancy in a dataset where every transmit sensitivity is modulated by every receive sensitivity is intuitively evident, as data with $N_{Tx}$ transmit channel and $N_{Rx}$ receive channels, results in $N_{Tx} \cdot N_{Rx}$ images, but only at most $N_{Tx} + N_{Rx}$ of them are linearly independent.

We investigated three different approaches for simultaneously reconstructing transmit and receive sensitivity maps in a single rank-constrained reconstruction. We formulated these all in a comprehensive multi-dimensional tensor model (k-space by transmit by receive), with the three approaches differing in how the low-rank constraints are imposed onto the tensor (via different matricizations or tensor-unfoldings).

The first method investigated (VC) is similar to the virtual coils approach in TIAMO(17), where the transmit and receive sensitivity combinations are used as virtual coils with the number of virtual coils equal to $N_{Tx} \cdot N_{Rx}$. The low-rank constraint was imposed on a tall matrix formed by the vertical concatenation of the block-Hankel matrix from each "virtual coil". However, this reconstruction rapidly deteriorated as the acceleration factor increased. This is likely because the matrix dimensionality is not well suited to the low rank constraint, as the matrix becomes significantly longer than it is wide (i.e there are more coils than elements in kernel). This is also reflected in the SNR performance where the virtual coils approach shows significantly worse RMSE with lower SNR.

The second method (PRIMO) uses the same transmit and receive channel concatenation scheme as PRIMO(10), with the low-rank constraint enforced on a matrix formed by vertically concatenating the block-Hankel matrices from each receive channel, and horizontally concatenating each transmit channel. This results in a more square matrix dimensionality than the VC approach, and a more effective low-rank representation. This approach performed significantly better than the VC method at lower acceleration factors (R≤4), but did exhibit similar or worse performance at high acceleration factors.

The third proposed method (TxLR) makes use of the symmetry in the transmit-receive array data by employing simultaneous low-rank constraints on two different unfoldings of the tensor data. One constraint is identical to the PRIMO constraint, while the second constraint enforces low-rankness on a matrix formed by horizontal receive concatenation and vertical transmit concatenation. The use of both these constraints is based on the recognition of the interchangeable nature of the transmit and receive sensitivities, where we demonstrated that each of these unfoldings individually have low-rank structure and perform similarly using an 8-channel transmit and 8-channel receive system. However, the two unfoldings capture *different* low-dimensional features in the tensor data, and are not trivially related through a transpose operation, and therefore leveraging the low-rankness of both of the unfoldings simultaneously constitutes a more powerful constraint, leading to dramatically improved reconstruction fidelity as assessed by RMSE.

In all the simulations using the synthetic body data, and the in vivo assessments in the body and the brain, the TxLR approach clearly performed better than the other methods, particularly at higher acceleration factors. At lower acceleration factors, RMSEs were similar to those produced by the PRIMO method, but dramatic differences were observed at acceleration factors of R=6 and R=8. The TxLR reconstructions demonstrated that in an 8-channel transmit system, acceleration factors of R=8 are quite feasible with low RMSE in the body (ranging from 0.1 in the best performing body dataset, to 0.2 in the worst case) and the brain (RMSE < 0.1 for all slices). These results indicate that relative transmit calibration can be achieved in the same amount of time or faster than B0 calibration. The reconstruction performance in the brain data suggested that the proposed approach may benefit from additional receive channels (32 in the brain compared to 8 in the body), although with only a single brain dataset this is difficult to determine.

Reconstruction hyperparameters (kernel size = [5,5], rank threshold = 50) were selected based on the synthetic dataset. These hyperparameters were able to robustly produce high-fidelity reconstructions in the in vivo body and brain datasets, without any tuning to the specific coil geometry or organ. In the body dataset, however, we did show that the chosen rank threshold of 50 was not always optimal, suggesting that data-specific hyperparameter tuning could lead to further reductions in RMSE. In addition, because of the non-convex problem formulation, stopping criteria for the iterative reconstruction can be important to avoid diverging from a local optimum. Here, we show that with knowledge of the channel-wise receiver noise characteristics, using a chi-square stopping heuristic resulted in near-optimal reconstruction performance.

In the body, B1+ mapping methods need to account for an increased dynamic range(22). Previous work has increased the dynamic range by using multiple maps at different excitation voltages(22). 3D methods have the potential to achieve this by improving the signal to noise efficiency and thus lower the noise floor, however for practical 3D mapping in the body acceleration is required.

Accelerations of 8 fold (matching the number of transmit states) worked best for an image matrix of 36x36, however a 24x24 matrix is sufficient to capture the dominant transmit modes in the heart (see Supporting Figures S2 and S3) where a six fold acceleration is reliable for the heart data. A 3D relative transmit mapping acquisition on an 8 channel system requires 4608 (24x24x8) lines of k-space data taking 16.1 s (TR=3.5ms) or 41 heart beats (cardiac window of 400 ms). An acceleration factor of 6 brings this down to 2.1s or only 7 heart beats for the whole torso. This becomes more significant when increasing the transmit channel count, for example mapping a 32 channel transmit system(23) would take 65 s (24x24x32 x 3.5ms), or take 161 heart beats, accelerating by a factor of 6 would bring this down to only 11 s, or 27 heart beats.

The transmit low rank (TxLR) approach could also be applied to absolute B1+ mapping. The 3D DREAM method proposed by Ehses(24) uses GRAPPA with a receive sensitivity pre-scan to accelerate the acquisition and reduce the echo train length. TxLR acceleration could be used to apply this method to parallel transmit mapping, removing the need for a pre-scan and enabling higher acceleration rates. Alternatively it could be used to accelerate methods like B1TIAMO(25) or interferometry methods(26) by accelerating the relative mapping part of the acquisition and applying these maps to reconstruct under sampled absolute maps.

Using TxLR in practice may be limited by the reconstruction time of the current implementation, although not prohibitively long, care would need to ensure further data can be acquired while reconstruction is in process. A rapid acquisition time will improve the robustness of the acquisition by reducing it susceptibility to motion and may further enable the characterisation of motion induced changes.

## 6. Conclusion

In conclusion calibrationless image reconstruction can be used to reconstruct transmit array sensitivity maps in highly accelerated conditions. Transmit low rank method harness redundancies in both transmit and receive coil profiles to enable reliable acceleration of transmit maps without the need for a calibration scan. The method enables acceleration factors of eight, equal to the number of transmit coils.

## Acknowledgements

This research was funded by the EPSRC (EP/T013133/1) and the Royal Academy of Engineering (RF201617\16\23). The Wellcome Centre for Integrative Neuroimaging is supported by core funding from the Wellcome Trust (203139/Z/16/Z). AH acknowledges support from the BHF Centre of Research Excellence, Oxford (RE/13/1/30181).

# Figure Captions

Figure 1 - Diagram showing how (a) a multi-dimensional k-space dataset with dimensions kx, ky, Rx and Tx can be transformed to (b) a 4th order tensor with block-Hankel frontal slices. (c-e) show different matricizations or unfoldings of the tensor, corresponding to: (c) the virtual coil matrix (V_C), (d) the PRIMO-style receive vertical concatenation matrix (R_C), and (e) the transmit vertical concatenation matrix (T_C). The proposed TxLR method uses low-rank constraints on both R_C and T_C unfoldings of the tensor.

Figure 2 - Singular value distributions for the T_C (a) and R_C (b) unfoldings of the synthetic data (blue). The shaded regions show the minimum and maximum range for the singular values across all slices in the dataset. In orange, singular value distributions for random matrices are shown for comparison.

Figure 3 - RMSE for the VC (blue), PRIMO (orange) and TxLR (yellow) methods evaluated in the synthetic body dataset. Shaded regions indicate the standard deviation of the RMSE across 48 slices.

Figure 4 - RMSE in the synthetic data at varying PSNR and acceleration factors. Each VC (blue), PRIMO (orange), and TxLR (yellow) scatter point represents the RSME for a single slice. Acceleration factors (a) R=4, and (b) R=8 are shown.

Figure 5 - Exploring the impact of various parameter choices on the TxLR approach. (a) RMSE at different kernel sizes and rank thresholds, using a fixed 50 iteration reconstruction; (b) similar to (a), but with a retrospectively selected number of iterations that produces the smallest RMSE; (c) line plot of the parameter space outlined in the red boxes of (a,b); (d) the result of varying the T_C and R_C constraints independently.

Figure 6 - (a) RMSE in the synthetic data at R=8, resulting from reconstructions using the T_C constraint (blue), R_C constraint (orange, PRIMO), and simultaneous T_C+R_C constraints (yellow, TxLR), with each marker corresponding to a single slice. (b) Reconstructed images and (c) transmit field maps corresponding to a representative Tx/Rx pair (Tx channel 4, Rx channel 3), at slice z=24. The number in the difference images correspond to RMSE.

Figure 7 - RMSE for different matrix sizes in the proposed method, at R=2, 4, 6 and 8. Each marker represents the RMSE from a single slice. Matrix sizes of 18x18 (blue), 24x24 (orange), 36x36 (yellow) and 48x48 (purple) were evaluated.

Figure 8 - RMSE vs iteration count for the proposed method, evaluated at PSNRs of 40 (yellow), 50 (orange) and 60 dB (blue) and acceleration factors of (a) R=2, (b) R=4, (c) R=6, and (d) R=8. The RMSE

at a fixed iteration count of 50 is denoted by the dashed black line, whereas the circles denote the number of iterations selected by the chi-square heuristic.

Figure 9 - Results for the in vivo 2D body data. (a) RMSE for the best performing subject and (b) the worst performing subject, showing the VC (blue), PRIMO (orange) and TxLR (yellow) methods against acceleration factors of R=2-8. (c) Estimated transmit sensitivity maps and the magnitude of the complex differences to the ground truth for Subject 4, Tx channel 3, at R=4 and (d) R=8. The values in each difference map correspond to the sensitivity map complex RMSE compared to ground truth.

Figure 10 - Results from the in vivo brain dataset. (a) RMSE at acceleration factors R=2, 4, 6 and 8, for the VC method (blue), PRIMO (orange) and TxLR (yellow), with markers representing the RMSE for each of the 36 slices. (b,c) Representative (Tx channel 4) transmit sensitivity maps, showing (b) magnitudes maps and magnitude differences, and (c) phase maps and phase difference images. The values inset on the difference maps in (b) correspond to the complex RMSE compared to ground truth

Figure S1 - RMSE for different iteration counts and different rank constraints. (a) VC method, (b) PRIMO, and (c) TxLR. (d) Plot of the RMSE vs iteration at rank 50.

Figure S2: RMS error between ESPIRiT reconstructed transmit sensitivity maps and a reference created with fully sampled transmit sensitivity data

Figure S3: For an ESPIRiT reconstruction with a 24 x 24 matrix and 6 x 6 kernel, shown is a) the difference to the fully sampled reference, and b) the RMS error (of all 8 combined).

Figure S4 - All 6 subjects from the in vivo body dataset. Bar lines denote the RMSE at the a priori rank threshold of 50, and the black dots denote the RMSE assuming an optimal choice of rank threshold.

# Supporting Information

Table S1: Relative transmit mapping acquisition protocols

|  | Brain – 3D | Body (1) – 2D | Body (2) – 2D | Body (3-6) – 2D |
|---|---|---|---|---|
| FOV (mm) | 220 x 220 x 163 | 400 x 400 x 8 | 300 x 300 x 10 | 360 x 360 x 8 |
| Matrix | 128 x 128 x 96 | 144 x 129 | 144 x 144 | 128 x 128 |
| Flip angle / reference voltage | 3° / 66 V | 4° / 200 V | 4° / 200 V | 7° / 200 V |
| Bandwidth (Hz/px) | 662 Hz/px | 694 Hz/px | 694 Hz/px | 797 Hz/px |
| TR (ms) | 6.0 ms | 3.6 ms | 3.5 ms | 3.2 ms |
| TE (ms) | 2.04 | 1.45 ms | 1.53 ms | 1.04 ms |
| Duration (s) | 688 s | 8 heart beats | | |

Figure S1 justifies the parameter choices made for the various reconstructions. All methods showed optimal RMSE around a rank threshold of 50. For the single constraint methods VC and PRIMO, optimality was achieved at the maximum evaluated iteration number of 100, whereas for the double constrained TxLR, optimality was achieved around iteration 50.

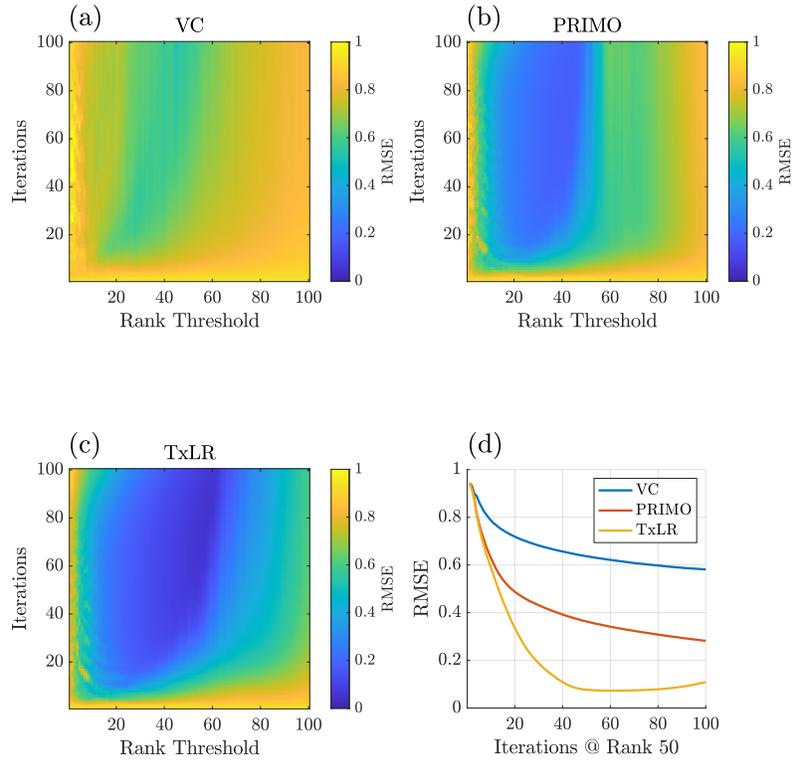

*Figure S1 - RMSE for different iteration counts and different rank constraints. (a) VC method, (b) PRIMO, and (c) TxLR. (d) Plot of the RMSE vs iteration at rank 50.*

## Justification for minimum measurement matrix for body relative transmit mapping

Using a single transverse slice from the middle of the synthetic body data set described in this work, the proton density was modulated by the transmit sensitives and Fourier transformed into k-space. Different measurement matrices around the centre of k-space were taken, including 10x10 and upto 48x48 in steps of 2. ESPIRiT(16) was used with varying kernel sizes of 3 to 10 to reconstruct transmit sensitivity maps. The RMS error in these reconstructed maps was compared to transmit sensitivity maps calculated from the full resolution data.

Figure S2 plots the RMS error for different kernel and matrix sizes, and figure S3 shows the error when using ESPIRiT to create a relative transmit sensitivity map.

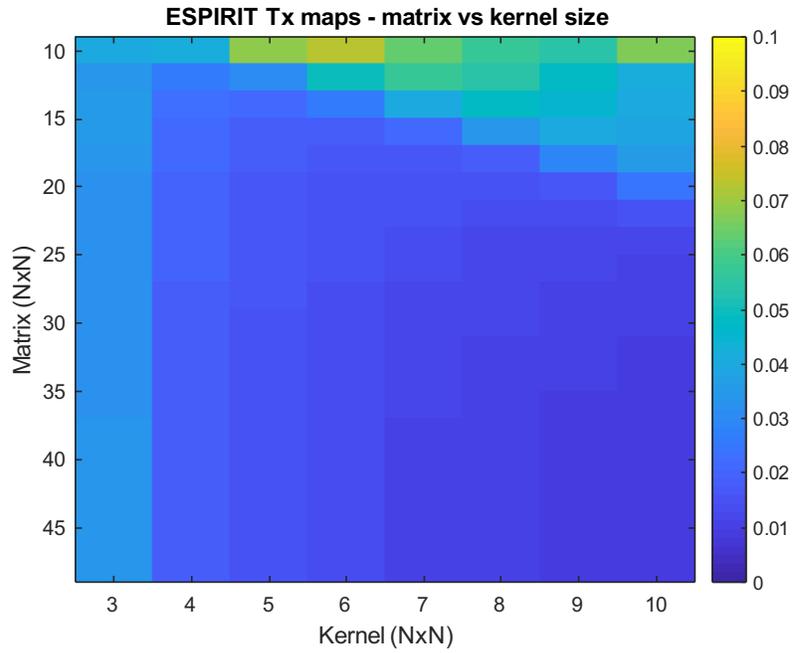

*Figure S2: RMS error between ESPIRiT reconstructed transmit sensitivity maps and a reference created with fully sampled transmit sensitivity data*

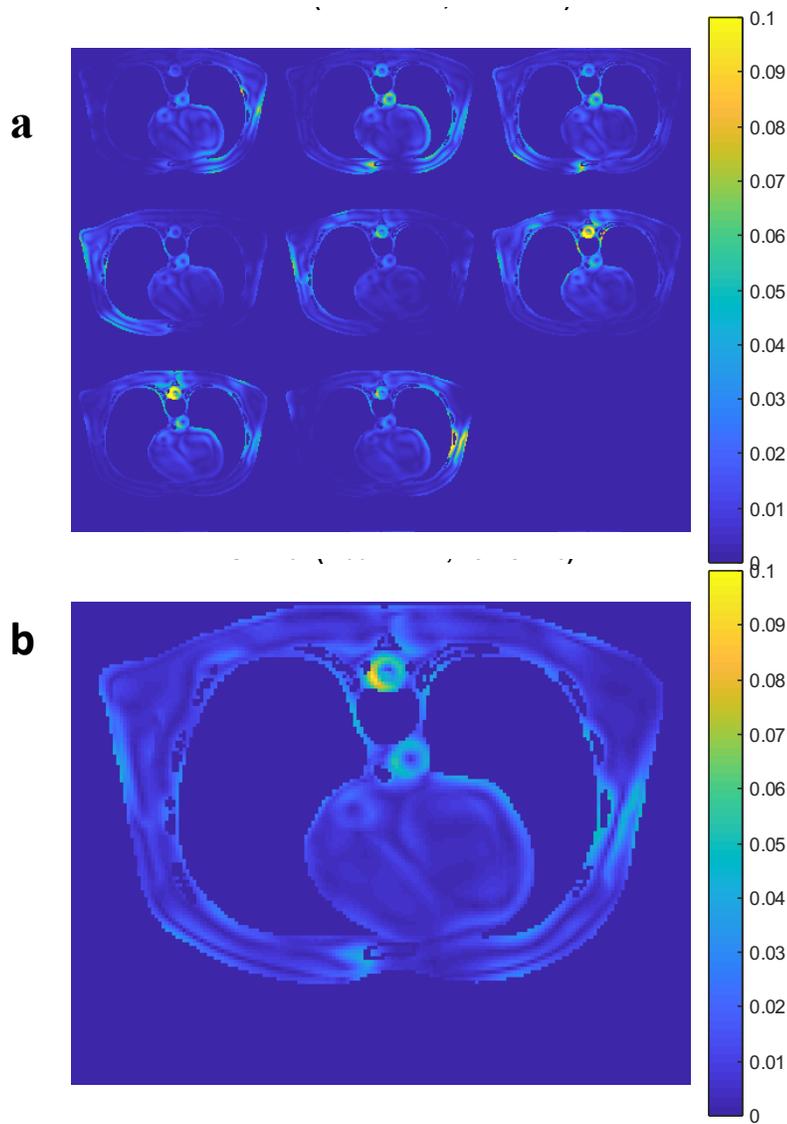

*Figure S3: For an ESPIRiT reconstruction with a 24 x 24 matrix and 6 x 6 kernel, shown is a) the difference to the fully sampled reference, and b) the RMS error (of all 8 combined).*

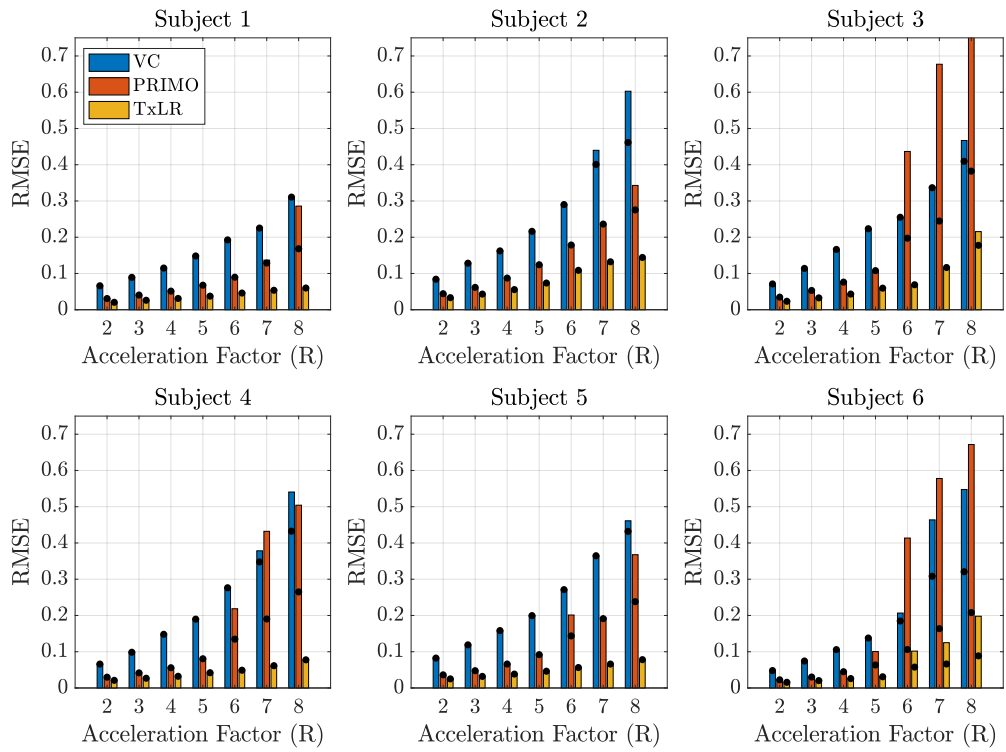

*Figure S4 - All 6 subjects from the in vivo body dataset. Bar lines denote the RMSE at the a priori rank threshold of 50, and the black dots denote the RMSE assuming an optimal choice of rank threshold.*